\documentclass[fleqn,10pt]{wlscirep}

\title{Cholesterol impairment contributes to neuroserpin aggregation}

\author[1,+]{Costanza Giampietro}
\author[2,+]{Maria Chiara Lionetti}
\author[3]{Giulio Costantini}
\author[2]{Federico Mutti}
\author[3, 4, 5, 6]{Stefano Zapperi}
\author[2,*]{Caterina A.M. La Porta}

\affil[1]{FIRC Institute of Molecular Oncology, 20139 Milan, Italy}
\affil[2]{Center for Complexity and Biosystems, Department of Biosciences, University of Milano, via Celoria 26, 20133 Milano, Italy}
\affil[3]{Center for Complexity and Biosystems, Department of Physics, University of Milano, via Celoria 16, 20133 Milano, Italy}
\affil[4]{CNR - Consiglio Nazionale delle Ricerche,  Istituto di Chimica della Materia Condensata e di Tecnologie per l'Energia, Via R. Cozzi 53, 20125 Milano, Italy}
\affil[5]{ISI Foundation, Via Alassio 11C, Torino, Italy}
\affil[6]{Department of Applied Physics, Aalto University, P.O. Box 14100, FIN-00076, Aalto, Finland}

\affil[*]{Corresponding author: caterina.laporta@unimi.it}

\affil[+]{these authors contributed equally to this work}

\begin{abstract}
Intraneural accumulation of misfolded proteins is a common feature of several neurodegenerative pathologies including Alzheimer's and Parkinson's diseases, and Familial Encephalopathy with Neuroserpin Inclusion Bodies (FENIB). FENIB is a rare disease due to a point mutation in neuroserpin which accelerates protein aggregation in the endoplasmic reticulum (ER). Here we show that cholesterol depletion induced either by prolonged exposure to statins or by inhibiting the sterol regulatory binding-element protein (SREBP) pathway also enhances aggregation of neuroserpin proteins. These findings can be explained considering a computational model of protein aggregation under non-equilibrium conditions, where a decrease in the rate of protein clearance improves aggregation. Decreasing cholesterol in cell membranes affects their biophysical properties, including their ability to form the vesicles needed for protein clearance, as we illustrate by a simple mathematical model.  Taken together, these results suggest that cholesterol reduction induces neuroserpin aggregation, even in absence of specific neuroserpin mutations. The new mechanism we uncover could be relevant also for other neurodegenerative diseases associated with protein aggregation.
\end{abstract}

\begin{document}

\flushbottom
\maketitle
%
%
\thispagestyle{empty}


\section*{Introduction}

Conformational diseases, such as Alzheimer’s and Parkinson’s diseases, spongiform encephalopathies and serpinopathies, are an increasingly common class of neurological disorders characterized by the aggregation of aberrant conformations of proteins. Familial Encephalopathy with Neuroserpin Inclusion Bodies (FENIB)~\cite{miranda2004,miranda2008} is an autosomal dominant inclusion body dementia characterized by protein aggregation within the  
endoplasmic reticulum (ER) ~\cite{miranda2004,miranda2008}. FENIB is associated with mutations in neuroserpin~\cite{roussel2011}, a protein belonging to the superfamily of serpins that plays an important role in brain development, neuronal survival, and synaptic plasticity ~\cite{yepes2000}. Indeed, mutants of neuroserpin in FENIB patients show accelerated rates of polymerization compared with wild type proteins, both at the protein level in vitro  ~\cite{belorgey2002,belorgey2004} and in cell models ~\cite{miranda2004,miranda2008}.  Moreover, neuroserpin is the most important inhibitor of the tissue plasminogen activator (tPA) in the brain which is known to be increased in Alzheimer's disease \cite{fabbro2011,subhadra2013,lee2015}.

Lipid metabolism has been shown to play a role in  FENIB, but also in other neurodegenerative diseases, from Alzheimer's to Parkinson's disease ~\cite{jones2010,pierrot2013,montag2012,roussel2013}. In FENIB the inhibition of hydroxymethyl glutaryl-CoA reductase (HMGCR) has a critical role in the clearance of mutant neuroserpin from the ER ~\cite{roussel2013}. Numerous studies have also reported that the modification of cholesterol content can affect amyloid precursor protein (APP) processing, which is needed for neuronal activity ~\cite{pierrot2013}. In this connection, it has been shown that the E693D (Osaka) mutation in APP promotes intracellular accumulation of Amyloid$\beta$ (A$\beta$), thus disturbing amyloid-mediated cholesterol efflux from the cell \cite{nomura2013}. Cholesterol has also been shown to influence APP processing and A$\beta$ generation by modulating A$\beta$- and Amyloid-$\gamma$-secretase activities  ~\cite{fassbender2001,grimm2008,refolo2001,yao2002}. In turn, APP cleavage products regulate cholesterol homeostasis ~\cite{green2013}.

In a recent paper, our group has shown by 3D numerical simulations and mean-field calculations that protein aggregation undergoes a non-equilibrium phase-transition controlled by the rates of protein synthesis and degradation, suggesting a crucial role of intracellular trafficking, particularly from the ER ~\cite{budrikis2014}. According to this study, a decrease in the rate of protein clearance from the ER can lead to rapid and irreversible protein aggregation in the ER itself \cite{budrikis2014}, leading to the onset of neurodegenerative diseases. 

Cholesterol plays a crucial role in regulating the properties of phospholipid membranes, affecting  their fluidity  and rigidity \cite{song1993bending,needham1990elastic,dimova2014recent}, the function and dynamics of membrane proteins
\cite{espenshade2002sterols,armstrong2012effect}, and thus vesicular trafficking within the cell \cite{wang2000cholesterol,Zhang2009,ridsdale2006cholesterol}. Hence, by regulating the biosynthesis of cholesterol it is possible to affect the form and function of all the membranes within the cell, including the ER.  In particular, experiments show that lowering the level of cholesterol delays ER-to-Golgi transport \cite{Runz2006,ridsdale2006cholesterol}. 

Cholesterol homeostasis is controlled by a family of transcription factors, known as sterol regulatory element binding proteins (SREBPs) ~\cite{bengoechea2007}. In cells with sufficient cholesterol supply, SREBPs are transmembrane proteins retained in the ER, associated with SREBP-cleavage-activating protein (SCAP), a cholesterol sensor ~\cite{feramisco2005}. Upon cellular cholesterol depletion, SREBP leaves the ER to reach the Golgi, where cleavage by site-1 protease (S1P) releases the amino-terminal half of SREBP, which can be further cleaved within its membrane-spanning helix by site-2 metalloproteinase (S2P) ~\cite{brown1999}. The mature processed form of SREBP is released in the cytosol and can translocate into the nucleus where it modulates the expression of several genes controlling cholesterol and fatty acid homeostasis ~\cite{horton2002}, including  HMGCR, HMG-CoA synthase (HMGCS), low density lipoprotein receptor (LDLR) and SREBP1/2 itself. SREBP1 activates fatty acid synthesis and SREBP2 cholesterol synthesis and uptake  ~\cite{goldstein2006}.

In this paper, we combine experiments and computational analysis to show that lowering the level of cholesterol either by using statins or by inhibiting SREBP1 and SREBP2 pathways with the small molecule betulin has a significant impact on the aggregation of non-mutated neuroserpin within the cell. Statins are competitive inhibitors of HMGCR, a rate limiting enzyme for cholesterol levels. They are widely used as plasma cholesterol lowering drugs in dyslipidemic patients. In spite of the clear effect of statins in lowering morbidity and mortality, in particular in cardiovascular events, many recent studies suggest that prolonged use of statins in Alzheimer's disease patients leads to mixed outcomes and could contribute to statin-adverse effects ~\cite{friedhoff2001,fassbender2002,simons2002,vega2003,doraiswamy2004}. Whereas current guidelines encourage an aggressive use of statins to achieve long-term cholesterol lowering in order to decrease cardiovascular events, the resulting effect on cell membranes is unknown. Here, we focus our experimental analysis on  non-mutated (wild type) neuroserpin and directly observe a dramatic increase in aggregates when cells are  chonically exposed to statins. To confirm the direct involvement in aggregation of lower cholesterol levels, we use betulin,
a pentacyclic triterpene commonly isolated from the birch bark, that specifically inhibits the maturation of SREBPs by enhancing interaction between SCAP and INSIG thus promoting SREBPs retention in the ER \cite{tang2011}. Also in this case, we observe lower cholesterol and
more neuroserpin aggregation. Our findings are consistent with a cholesterol induced decrease in the protein clearance rate, leading to faster aggregation in our model \cite{budrikis2014}. To elucidate the biophysical mechanism underlying this observation, we study a mathematical model of coated vesicle release from lipid membranes, illustrating the crucial role of cholesterol for vesicle formation and therefore the impact of its unbalance on intracellular trafficking. Taken together, our results suggest that long-term treatment with statins may affect intracellular trafficking in a way to enhance protein aggregation.  

\section*{Results}

\subsection*{Prolonged treatment with statins enhances neuroserpin aggregation in HeLa cells}
\begin{figure}[htb] \centering 
\includegraphics[width=15cm]{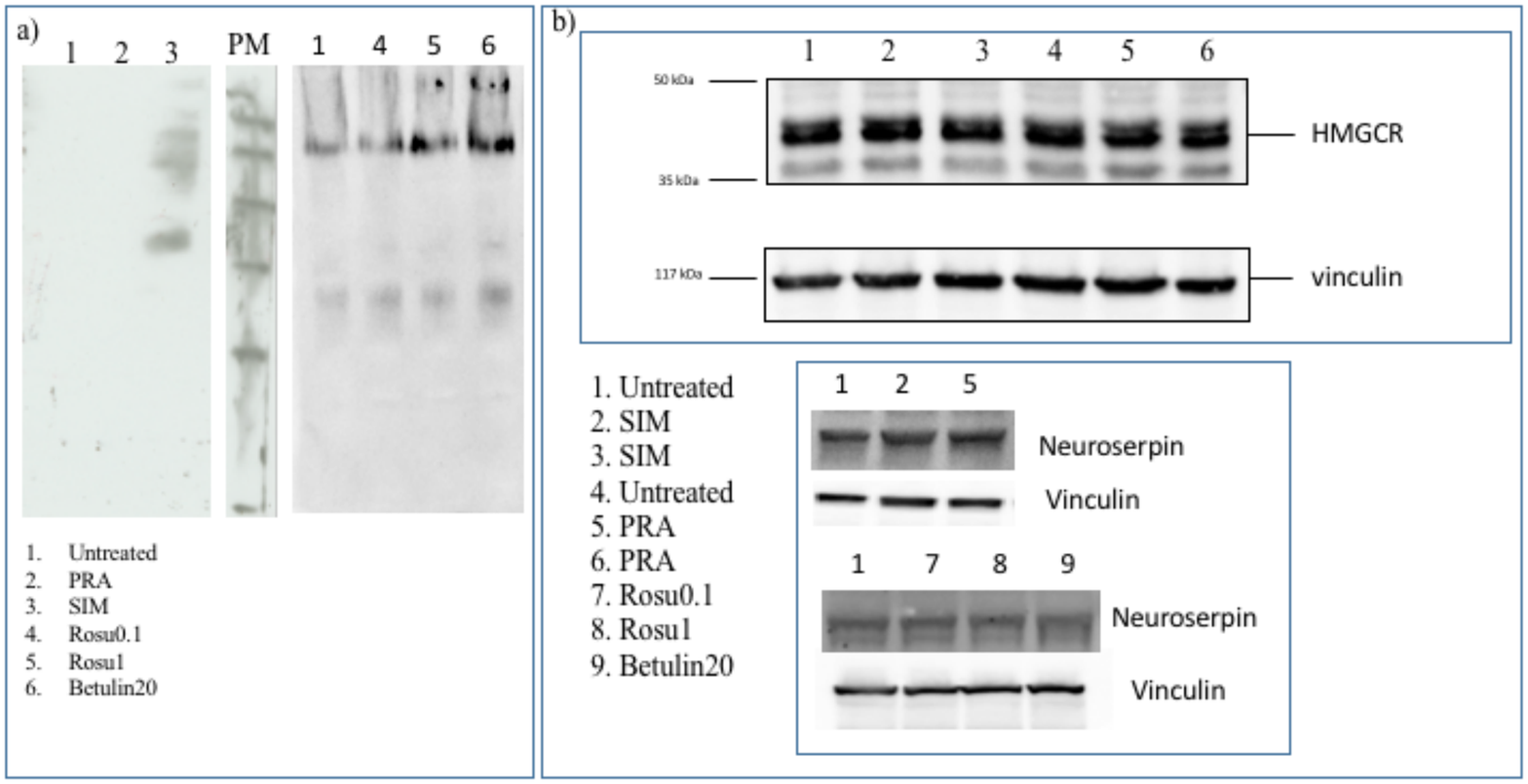}
 \caption{\label{fig:1} {\bf Aggregation of neuroserpin with statins or betulin} a) 100$\mu$g protein of HeLa cells or treated with 10$\mu$M PRA or 5$\mu$M SIM  or 0.1$\mu$M rosuvastatin or 1$\mu$M rosuvastatin or 20$\mu$M betulin  were loaded on 10\% native gel and transferred on PVDF membrane. The sheet was incubated with anti-neuroserpin (ab32901, Abcam, 1:500) overnight at 4$^\circ$C and  then with the secondary antibody anti goat-HRP (1:5000, ECL Blotting reagents (GE Healthcare RPN2109) / SuperSignal West Femto Maximum Sensitivity Substrate, Thermo scientific) for 1h at room temperature. Ponceau S solution (P7170, Sigma) was used to verify the correct loading of the samples. b) Effect of statins or betulin treatments on the level of expression of neuroserpin or HMGCR. 10$\mu$g total protein were loaded on 10\% polyacrilamide gel, transferred on PVDF and incubated with anti-neuroserpin (ab32901, AbCAM, 1:500)  or with anti-HMGCR (1:1000, MAB5374, Millipore) overnight at 4$^\circ$C). Vinculin (1:10000, Sigma) was used as housekeeping. }
 \end{figure}

\begin{figure}[htb] \centering 
\includegraphics[width=13cm]{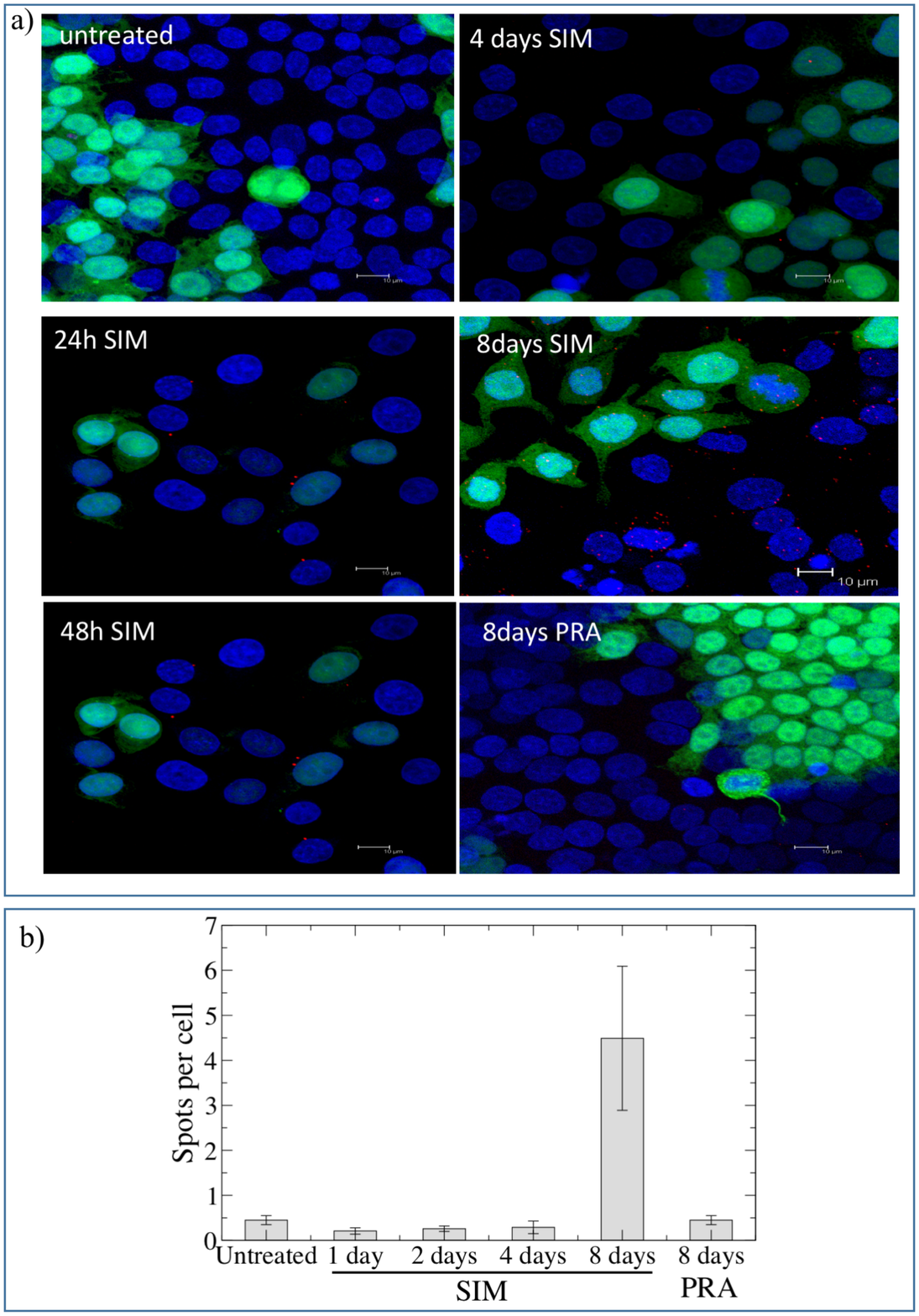}
 \caption{\label{fig:2}  {\bf Duolink In situ Staining of neuroserpin in HeLa untretaed or treated with SIM or PRA.} a) Hela-GFP and Hela cells were mixed 1:1 and treated with 5$\mu$M SIM or 10$\mu$M PRA  up to 8 days, fixed in 3.7\% paraformaldeide and incubated with the primary antibody anti-neuroserpin (Ab32901, Abcam) coniugated PLA probes (1:50), overnight at 4$^\circ$C. The presence of aggregates are visualized using Duolink in Situ staining according to the manufacturer’s instructions. Nuclei are stained with DAPI. Images are acquired by Leica SP2  laser scanning confocal microscope. Each red spot represents a neuroserpin aggregate.  b) The number of spots per cell for each condition reported in panel a is quantified as described in the 
methods section.}
 \end{figure}

Simvastatin (SIM), pravastatin (PRA) and rosuvastatin are all known to be inhibitors of cholesterol synthesis by acting on HMGCR. PRA is, however, reported to be less effective both in patients and in vitro due to its hydrophylicity ~\cite{contermans1995}. We test the effect of long-term (i.e. up to 8 days) treatment with statins (5$\mu$M SIM or 10$\mu$M PRA or  0.1$\mu$M rosuvastatin or 1$\mu$M rosuvastatin) on the capability of neuroserpin to aggregate in HeLa cells. Fig.~\ref{fig:1}a shows that, in native gel, the treatment with SIM is able to induce multiple bands of neuroserpin, suggesting the presence of aggregates. We obtain similar results for the more powerful drug rosuvastatin, as reported in Fig.~\ref{fig:1}b ~\cite{mckenney2003}. In contrast, the exposure to PRA does not affect neuroserpin aggregation, as shown in Fig.~\ref{fig:1}a. To directly confirm the presence of neuroserpin aggregates after prolonged treatment with statins, we use the Duolink in Situ staining. 
Using this technique allows the detection of a fluorescent spot if at least two monomers of neuroserpin protein are in close proximity, such as when they form a dimer. We mix (1:1) HeLa with HeLa-GFP cells to confirm that possible aggregation is independent from the specific cell line used. Moreover, in HeLa---GFP cells the cytoplasm is green. We then treat the cells with statins for up to 8 days.  Long-time treatment (8 days) with 5$\mu$M SIM or 0.1$\mu$M rosuvastatin induces neuroserpin aggregation while shorter incubation times with 5$\mu$M SIM or 8 days of treatment with 10$\mu$M PRA do not induce any effect, as shown in Fig.~\ref{fig:2}. Moreover, 8 days of treatment with 0.1$\mu$M rosuvastatin gives a stronger effect with respect to 5$\mu$M SIM  Fig.~\ref{fig:2} and Fig.~\ref{fig:4}.

\subsection*{Effect of the treatment with SIM, rosuvastatin or PRA on cholesterol and HMGCR levels in HeLa cells}
\begin{figure}[htb] \centering 
\includegraphics[width=15cm]{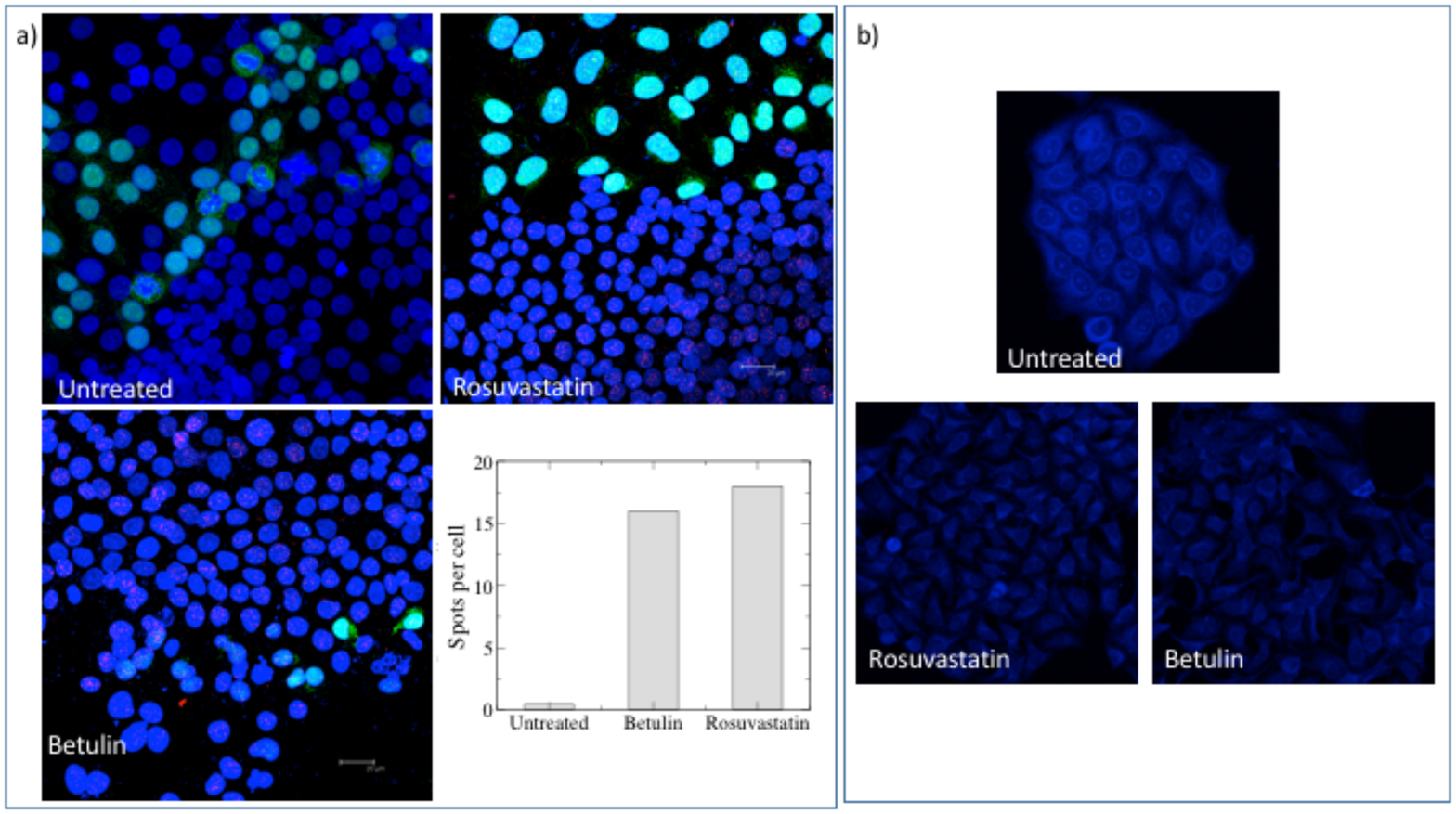}
 \caption{\label{fig:3} {\bf Detection of cholesterol with TNM-AMCA in untreated and treated HeLa cells after statins treatments.} Untreated or 5$\mu$M SIM or 10$\mu$M PRA treated HeLa cells for the time reported in the figure, were fixed with 3.7\% paraformaldeide and then incubated with TNM-AMCA (1$\mu$M) to stain cholesterol for 1h at room temperature. Images were acquired by Leica SP2 laser scanning confocal microscope.}
 \end{figure}

Filipin is commonly used to detect cholesterol, but it is not very stable, rapidly photobleaches and results in acute toxicity for living cells\cite{brajtburg1974}. To avoid these problems, we use a fluorescently labeled theonellamide (TNM-AMCA). TNM-AMCA was originally isolated from a marine sponge, and has been reported to bind in a specific manner to the 3$\beta$-hydroxyl group of cholesterol without acute toxicity ~\cite{nishimura2010,nishimura2013,arita2015}. 5$\mu$M SIM, 0.1$\mu$M rosuvastatin or 20$\mu$M betulin affect the level of cholesterol after 8 days of treatment (Fig.~\ref{fig:3} and Fig.~\ref{fig:4}), while PRA is less effective and does not change significantly the level of cholesterol\cite{contermans1995} (Fig.~\ref{fig:3}). Finally we investigate by western blot whether the treatment with statins affects the expression level of the key enzyme of cholesterol synthesis, HMGCR. As shown in Fig.~\ref{fig:1}c, no significant changes in the level of expression of this enzyme occurs after long-time treatment either with SIM or PRA.

\subsection*{Effect of the treatment with SIM, rosuvastatin or PRA on the level of expression of SREBP1 and SREBP2}
\begin{figure}[htb] \centering 
\includegraphics[width=15cm]{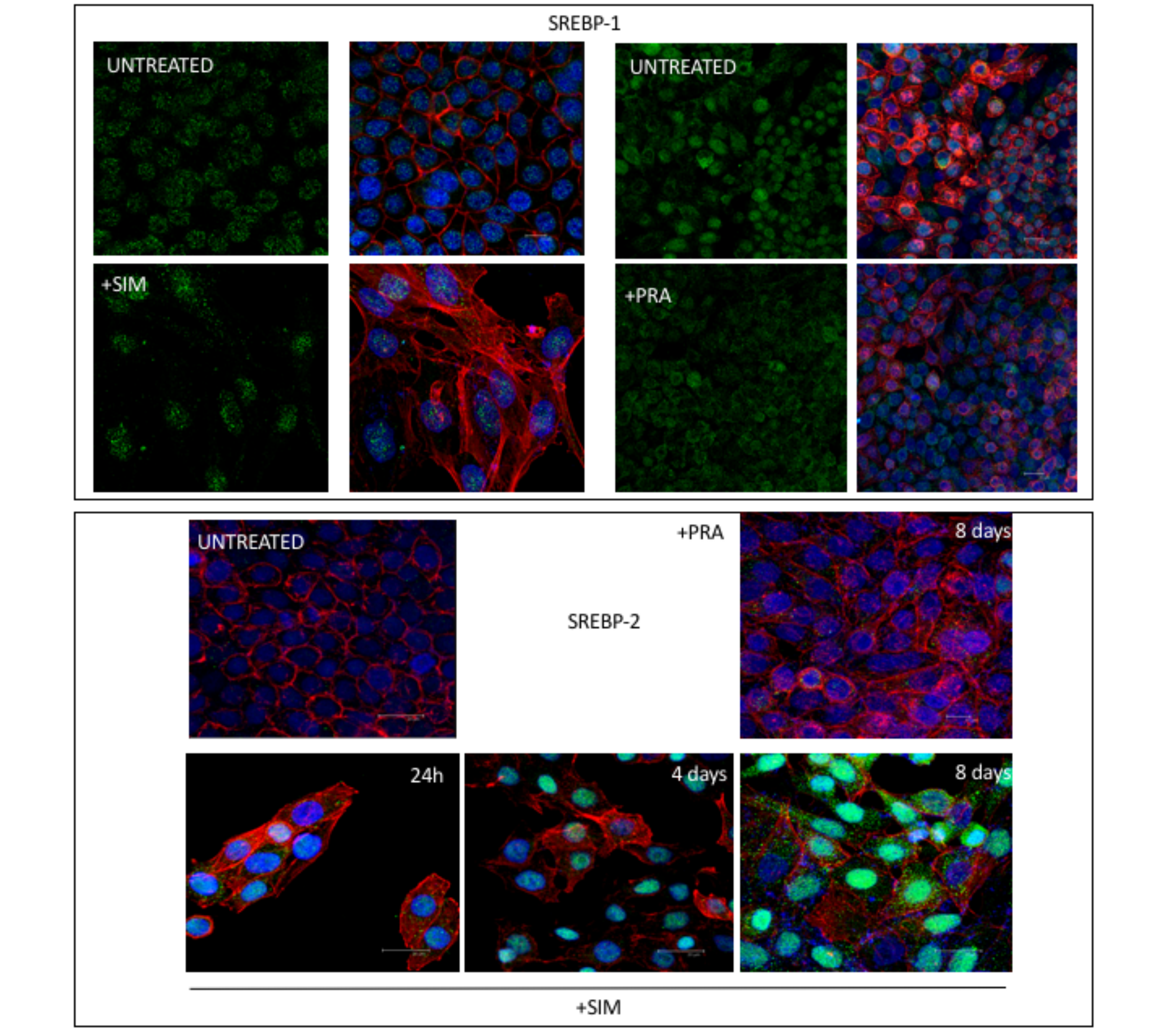}
 \caption{\label{fig:5}  {\bf Level of expression of SREBP1 and SREBP2 after SIM or PRA treatment in Hela cells.} Subconfluent cells are fixed with 3.7\%paraformaldeide after 8 days treatment with statins and incubated with anti-SREBP1 (1:100, ab28481) or anti SREBP-2 (1$\mu$g/ml ab30682, Abcam) overnight at 
 4$^\circ$C. Next, the cells are incubated with anti-rabbit Alexa Fluo488 for 1h at room temperature. Actin is stained with 1$\mu$g/ml Actistain-555 phalloidin for 1h at room temperature. Nuclei are stained with DAPI. Untreated cells are also shown for reference. Images are acquired by Leica SP2  laser scanning confocal microscope.}
  \end{figure}

\begin{figure}[htb] \centering 
\includegraphics[width=15cm]{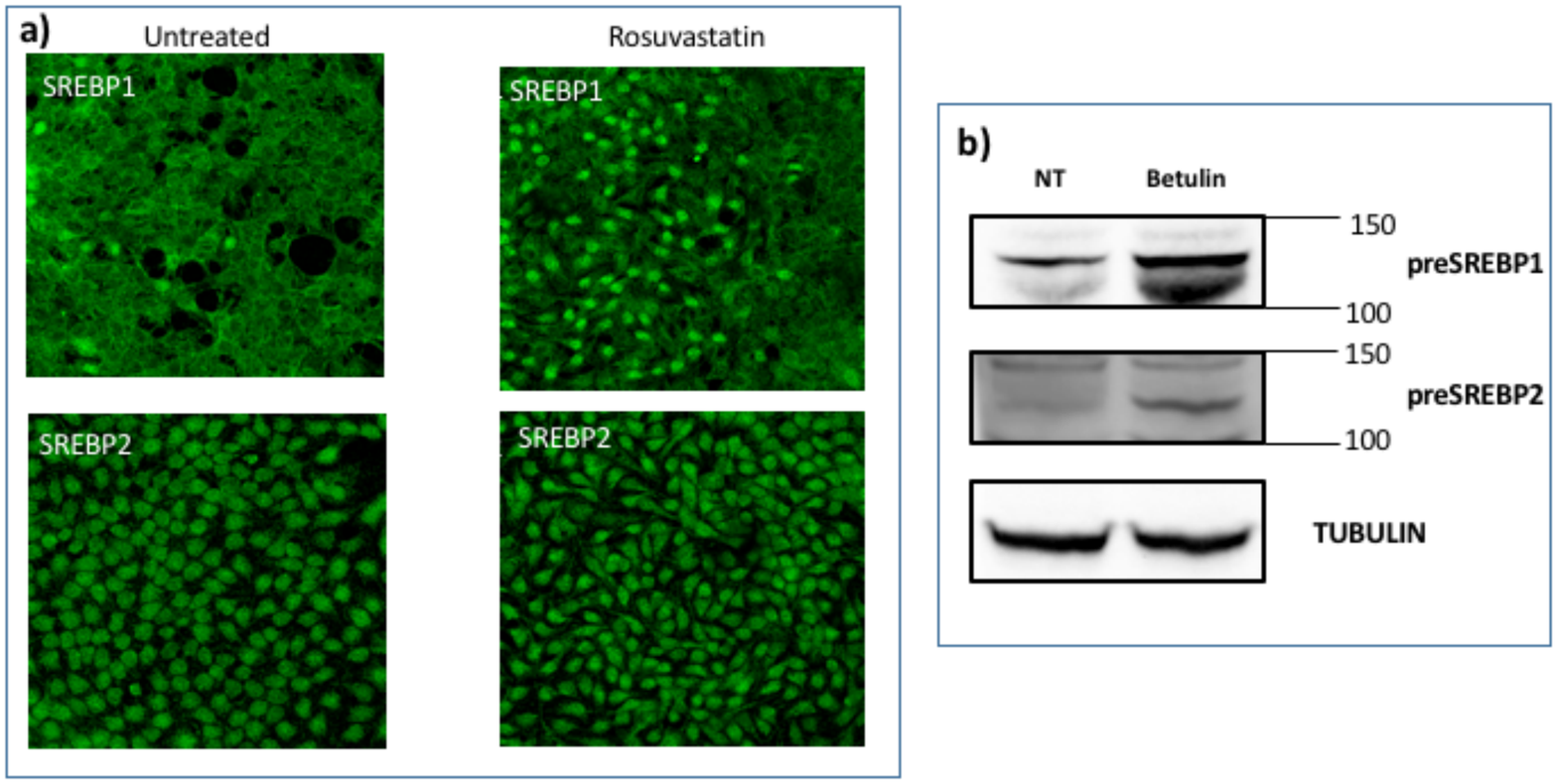}
 \caption{\label{fig:6}  {\bf a) Level of expression of SREBP1 and SREBP2 after rosuvastatin or betulin treatment in Hela cells.} Subconfluent cells a fixed with 3.7\%paraforldeide after 8 days of rosovastatin  (0.1$\mu$ M) or betulin (20$\mu$M) treatment and incubated with anti-SREBP1 ( 1:100, ab28481) or anti SREBP-2 (1$\mu$g/ml ab30682, Abcam) overnight at 4$^\circ$ C. Then, the cells are incubated with anti-rabbit AlexaFluo488 for 1h at room temperature. Nuclei are stained with DAPI. Untreated cells are also shown for reference. Images are acquired by Leica SP2  laser scanning confocal microscope. b) Western blot of SREBP1 and SREBP2 in Hela cells untreated and after 8 days of treatment with (20$\mu$M betulin. 10$\mu$g total protein were loaded on 10\% polyacrilamide gel, transferred on PVDF and incubated or anti-SREPB1 (1:500, ab28481, AbCam) or anti-SREBP2 (1:100, ab30682, AbCam) were used overnight at 4$^\circ$C. Mouse anti-$alpha$ Tubulin antibody (1:5000, Sigma) for 1h at room temperature was used as housekeeping. A secondary antibody anti goat-HRP (1:5000, ECL Blotting reagents (GE Healthcare RPN2109) / SuperSignal\texttrademark West Femto Maximum Sensitivity Substrate, Thermo scientific) was used for 1h at room temperature to detect chemiluminiscence. The bands reported are the precursor of SREBP1 and SREBP2, pre-SREBP}
 \end{figure}

SREBPs are transmembrane proteins expressed at the level of ER, associated with SCAP which is a cholesterol sensor ~\cite{feramisco2005}. Intracellular trafficking of SREBPs is sensitive to cholesterol depletion:  when the level of cholesterol is reduced, SREBP leaves the ER to reach the Golgi where it is cleaved. Then, the mature processed form of SREBP is released in the cytosol and translocates into the nucleus where it modulates the expression of several genes controlling cholesterol and fatty acid homeostasis\cite{horton2002} and SREBP1/2 itself. Furthermore, it is known that SREBP1 regulates fatty acid synthesis and SREBP2 cholesterol synthesis and uptake  ~\cite{goldstein2006}. We thus investigate by immunofluorescence the effect of statins on the level of expression of SREBP1 and SREBP2. Fig.~\ref{fig:5} shows that  SREBP1 is not affected by the either SIM or PRA treatments and SREBP2 increases significantly after 8 days of treatment with SIM.  Rosuvastatin, in contrast, affects mainly SREBP1 (Fig.~\ref{fig:6}a).

\subsection*{Effect of the treatment with betulin on the level of expression of cholesterol, neuroserpin aggregation, SREBP1 and SREBP2}
\begin{figure}[htb] \centering 
\includegraphics[width=15cm]{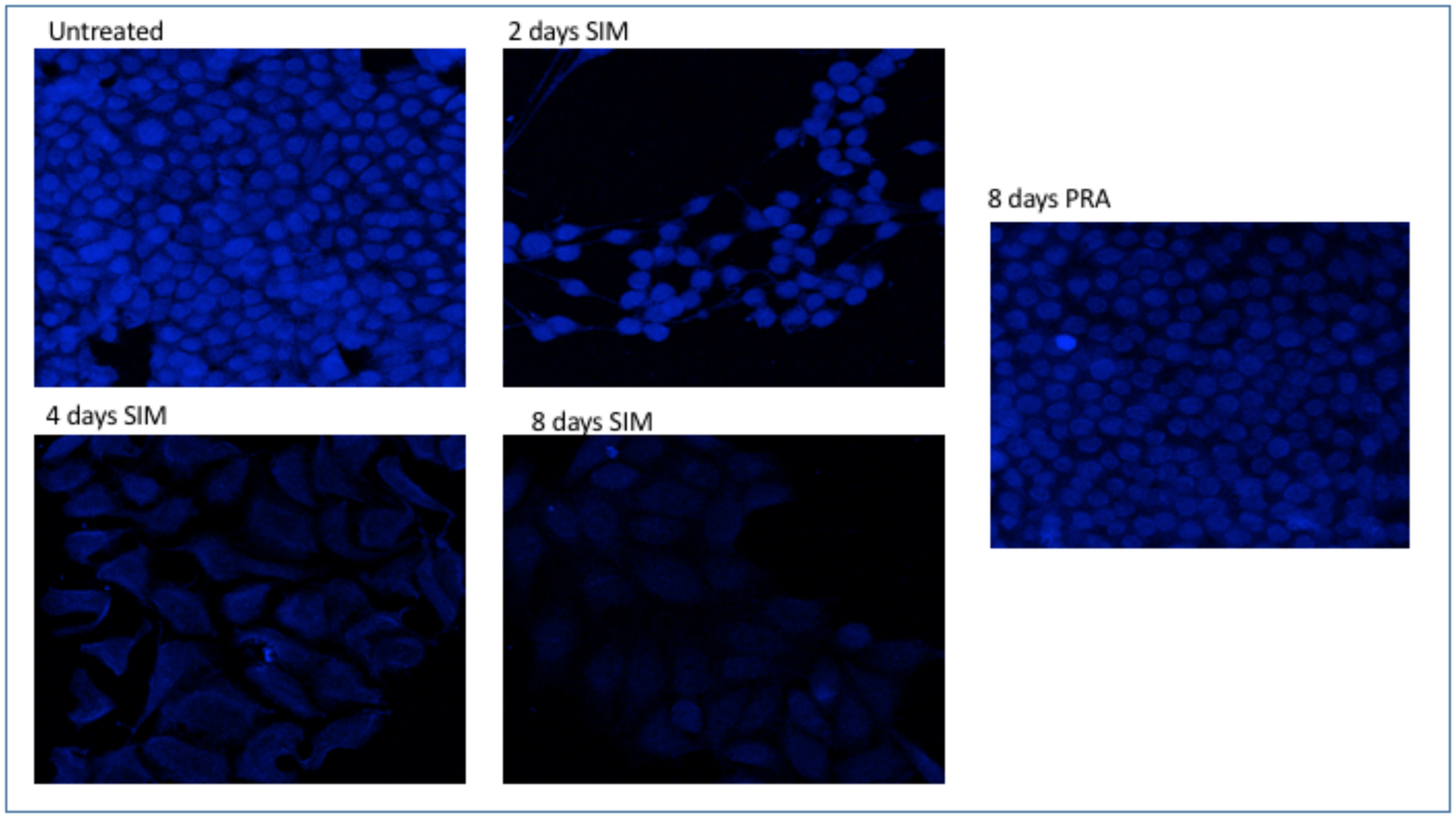}
 \caption{\label{fig:4}  {\bf Effect of rosuvastatin and betulin on neuroserpin aggregation   and cholesterol level}.  a) Hela-GFP and Hela cells were mixed 1:1, treated with 0.1$\mu$M rosuvastatin or 20$\mu$M betulin for 8days, fixed in 3.7\% paraformaldeide and incubated with the primary antibody anti-neuroserpin (Ab32901, Abcam) coniugated PLA probes (1:50), overnight at 4$^\circ$C. The presence of aggregates are visualized using Duolink in Situ staining according to the manufacturer’s instructions. Nuclei are stained with DAPI. Images are acquired by Leica SP2  laser scanning confocal microscope. Each red spot represents a neuroserpin aggregate. The number of spots per cell for each condition in the images showed in this panel is quantified as described in the methods section. b) Untreated or treated cells with  0.1$\mu$M rosuvastatin or 20$\mu$M betulin (8 days) were fixed with 3.7\% paraformaldeide and then incubated with TNM-AMCA (1$\mu$M) to stain cholesterol for 1h at room temperature. Images were acquired by Leica SP2 laser scanning confocal microscope.}
 \end{figure}

An alternative route to interfere with the cholesterol level and study its effect on neuroserpin aggregation is to inhibit  SREBPs activity using betulin, a small molecule that is known to inhibit the maturation of SREBP by inducing interaction of SCAP and Insig  \cite{soyal2015,tang2011}. Betulin induces a decrease of cholesterol after 8 days of treatment (Fig.~\ref{fig:4}) \cite{tang2011}  and an accumulation of the precursor of SREBPs (pre-SREBP), in agreement with Ref. \cite{krycer2012} \cite{quan2013} (Fig.~\ref{fig:6}b). Under these conditions, we show that the treatment for 8 days with 20$\mu$M betulin induces neuroserpin aggregation (Fig.~\ref{fig:1} and Fig. \ref{fig:4}a).     

\subsection*{Computational simulations reveal aggregation induced by the impairment of intracellular trafficking} 
\begin{figure}[htb] \centering 
\includegraphics[width=10cm]{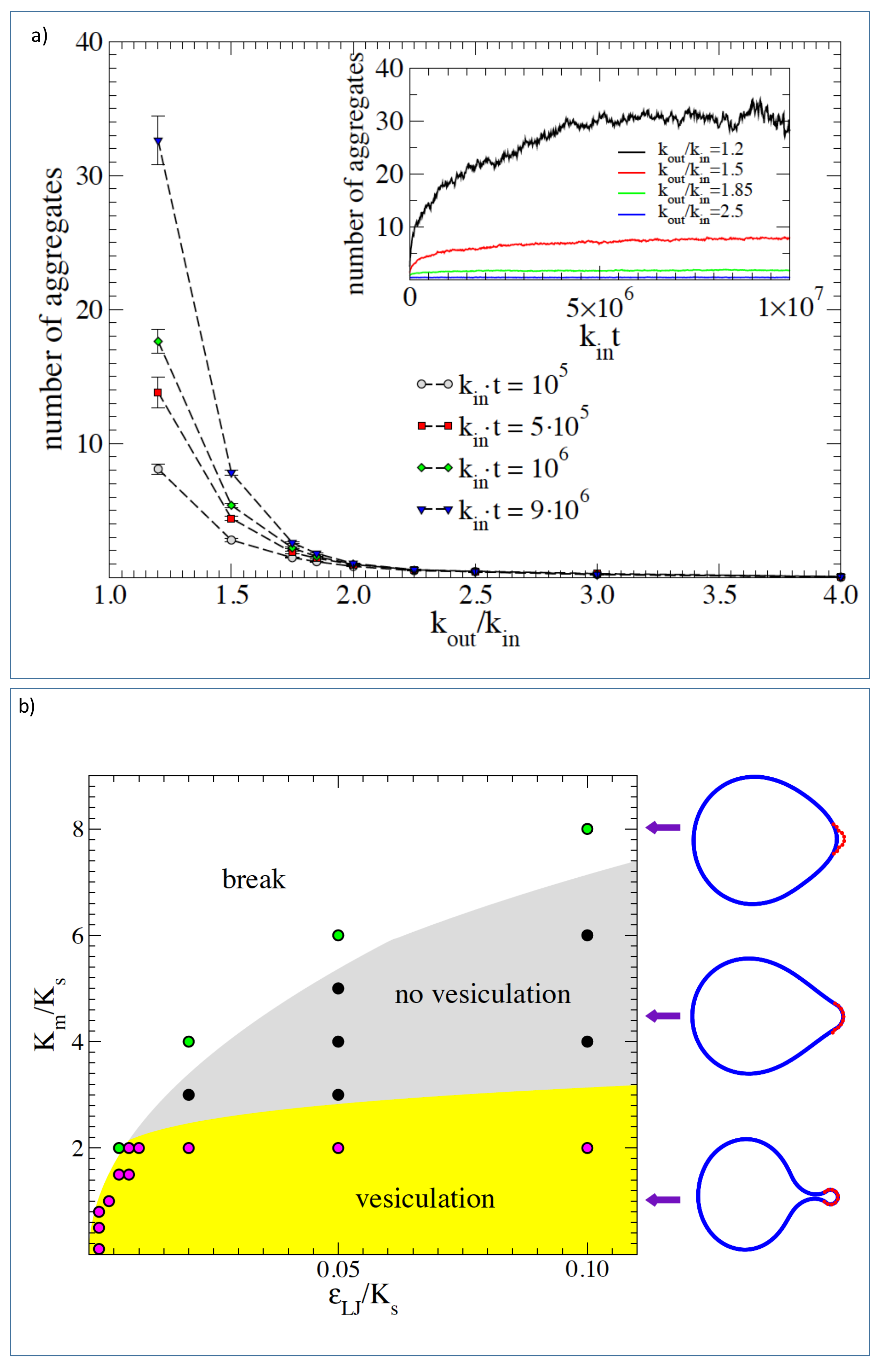}
 \caption{\label{fig:7}  {\bf Numerical simulations show protein aggregation induced by impairment of trafficking.} a) The main figure shows the average number of aggregates observed at a fixed time as a function of $k_{\rm out}/k_{\rm in}$, the ratio between rate of exit from the ER by means of vesicular transport and the rate of protein production. A reduced value of $k_{\rm out}$ leads to the formation of aggregates. The inset shows the time dependence of the average number of aggregates for different values of  $k_{\rm out}$. b) The phase diagram of the vesicle formation model. The stiffness ratio between the membrane and the external scaffold $K_m/K_s$ as a function of the intensity of the mutual interaction with respect to the scaffold rigidity $\epsilon_{LJ}/K_s$. For soft membranes there is almost always vesicle formation (yellow zone) or the breaking of the coat-membrane interface, if binding is not strong enough (white zone). Increasing the membrane stiffness leads to a new phase appears where the coat-membrane interface is stable but vesicle can not form (grey zone). 
Typical configurations in the different phases are reported on the right-hand side of the panel.}
 \end{figure}

In a previous paper our group developed a model three dimensional to explain the critical role for protein aggregation of the non-equilibrium conditions in the ER\cite{budrikis2014}. Here we implement the model in order to  understand how the impairment of intracellular trafficking might induce protein aggregation. 
The model performs coarse---grained simulations of diffusing monomers and polymers in a closed system, representing the ER, under a constant rate of protein production $k_{\rm in}$. Proteins exit the ER with a rate $k_{\rm out}$ when they are close to the boundaries simulating vesicular trafficking. Hence, the parameter $k_{\rm out}$ encapsulates the efficiency of intracellular trafficking which can be affected by the level of cholesterol in the cell. Fig.~\ref{fig:7}a shows that reducing $k_{\rm out}$, which should simulate an impairment in the intracellular trafficking leads to an increase in protein aggregation, thus recapitulating {\it in silico} the experimental observations.

\subsection*{Mathematical model illustrates mechanical regulation of vesicle formation} 
The mechanism by which the level of cholesterol determines the value of $k_{\rm out}$ is related to its role in affecting the biophysical properties of membranes \cite{song1993bending,needham1990elastic,dimova2014recent,espenshade2002sterols,armstrong2012effect} and thus their ability to form the vesicles needed for intracellular trafficking \cite{wang2000cholesterol,Zhang2009,ridsdale2006cholesterol}. 
Vesicle formation in the ER is assisted by the Coat protein complex II (COPII) that is recruited on the surface
of the ER membrane \cite{Kuehn1998}. The COPII complex self-assembles into a cage structure that encapsulate the lipid membrane forming a vesicle \cite{Zanetti2013}. The success of the process crucially depends on the binding strength of the COPII to the membrane and on its bending stiffness, as we can illustrate by a simple mathematical model. We  considers an elastic membrane coupled to a curvature induced coat, representing the action of a COPII cage in the ER. The model, described in details in the methods section includes two key dimensionless parameters: i) the ratio between the bending 
stiffness of the membrane $K_m$ and the one of the coating scaffold $K_s$, and ii) the ratio between the scaffold bending energy and its interaction energy with the membrane $\epsilon_{LJ}$. Simulations show that vesicle can only form is $K_m/K_s$ is small and $\epsilon_{LJ}/K_s$ is sufficiently large. For larger values of  $K_m/K_s$, the coating scaffold is not able to bend the membrane into a vesicle while if $\epsilon_{LJ}/K_s$ is small the coat detaches from the membrane. This behaviour is summarized in the phase diagram reported in Fig. \ref{fig:7}b. A reduction in the level of cholesterol in the ER membrane affects both $K_m$ \cite{song1993bending,needham1990elastic,dimova2014recent} and $\epsilon_{LJ}$ \cite{Runz2006}, thus displacing the system in different regions of the phase diagram. In particular, experiments
show that cholesterol depletion leads to a reduced recruitment of the Sec23p component of the COPII complex on the ER membrane, delaying ER-to-Golgi transport \cite{Runz2006}.

\section*{Discussion}

Long-term lowering serum low density lipoprotein cholesterol with statin drugs is used extensively and has proven very effective to reduce the incidence of cardiovascular events. De-novo cholesterol which is the target of statin therapy is found in all membranes and lipid based bodies, contributing to vesicle formation and migration, as well as other membrane functions ~\cite{ikonen2008}. We can therefore expect that regulating the biosynthesis of cholesterol could change the form and function of every membrane within the cell. For this reason, statin therapies could also cause potential harm~\cite{kiortsis2007}. The role of ER in the aggregation of misfolded proteins, such as neuroserpin, and therefore as a crucial determinant of cellular toxicity, is important in many neurodegenerative diseases. Our group recently demonstrated the presence of a phase transition towards rapid protein aggregation due to a breakdown of homeostasis in intracellular organelles, such as the ER, controlled by the rates of protein synthesis and clearance, $k_{\rm in}$ and $k_{\rm out}$, respectively ~\cite{budrikis2014}. The open question stated in the conclusions of our earlier paper was related to the identification of possible biological processes and factors that would tune these key parameters into an aggregation-prone pathological phase ~\cite{budrikis2014}. 
In the present paper, we combine experiments and computational models to investigate this critical issue. 

We focus our attention on cholesterol depletion by statins as a potential inhibitor protein clearance from the ER. We investigate three different statins in a chonic treatement (8 days): SIM, rosuvastatin and PRA. Rosuvastatin has been shown to be more effective than SIM ~\cite{mckenney2003}, while PRA is less 
effective when compared with SIM~\cite{contermans1995}. Here we show indeed that long-term treatment with SIM or rosuvastatin decreases the level of cholesterol, increases the level of expression of SREBP,  compensating for cholesterol depletion, and finally enhances neuroserpin aggregation even in the absence
of specific mutations. In contrast, PRA is less effective than the others two statins and therefore does not affect neither the level of cholesterol nor neuroserpin aggregation  ~\cite{contermans1995}. 

To ensure that the aggregation we observe for neuroserpin is  really due to the lowering of cholesterol in the cell rather to a direct interaction between statins and neuroserpin, we interfere with the cholesterol level indirectly by targeting SREBP  through betulin, whose action ultimately leads to cholesterol depletion \cite{tang2011}. We found that betulin decreases the level of cholesterol and  accumulates pre-SREBPs \cite{tang2011,quan2013,krycer2012}. Also in this case, we find an enhancement of neuroserpin aggregation after long term treatment (8 days).

It is well known that N-glycosylation plays determinant roles in protein folding and trafficking, and N-glycosylated proteins are especially important in regulating extracellular activities. Recent papers show that aberrant N-glycosilation happens frequently in relation to human diseases, including Alzheimer's\cite{wang1996} and FENIB, as pointed out in a recent paper \cite{ moriconi2015}. Furthermore, a systematic and quantitative analysis of surface proteins was carried out in HepG2 liver cells treated with statin showing that many glycosilation sites are downregulated compared to untreated cells ~\cite{xiao2016}. On the other hand, statins  inhibit the pathway of dilichol,  which plays a critical role in protein N-glycosilation and acts as a membrane anchor for the formation of a precursor oligosaccharide ~\cite{burda1999}. Therefore, the long-treatment with statins could lead to important impairment of intracellular trafficking, effectively decreasing the clearance rate $k_{\rm out}$  ~\cite{budrikis2014}. This is shown directly by numerical simulations of a protein aggregation model where reducing $k_{\rm out}$ leads to faster aggregation.

The role of cholesterol on the biophysical properties of lipid membranes has been the subject of extensive experimental investigation. It is known that the level of cholesterol affects  the membrane bending stiffness $K_m$ \cite{song1993bending,needham1990elastic,dimova2014recent}  the dynamics of membrane proteins \cite{espenshade2002sterols,armstrong2012effect} and has thus an effect on vesicle formation and trafficking \cite{wang2000cholesterol,Zhang2009,ridsdale2006cholesterol}. In particular, experiments show that the attachment of coating proteins, such as COPII, to membranes is dependendent on cholesterol \cite{Runz2006}. Here, we develop simple mathematical model that can incorporate into a single framework all this experimental observations and help predict the effect of statins on the intracellular trafficking. While the model is too simple to provide a quantitative  explanation of vesicle formation in dependence on the cholesterol level, the phase diagram illustrates the possibility that small changes of key membrane biophysical parameters such as the bending stiffness or the binding affinity of coat proteins, all known to be affected by cholesterol, result in a net impairment of vesicle formation.

Recent results in the literature show an increased level of neuroserpin in Alzheimer's disease patients, the critical role of tPA and the association between neuroserpin and A$\beta$ plaques in Alzheimer's brain tissues \cite{fabbro2009}. Our paper shows that neurosperpin aggregation can be modulated by environmental factors affecting intracellular trafficking, even in the absence of deleterious mutations that are known to induce aggregation \cite{belorgey2002,belorgey2004,miranda2004,miranda2008}. The aggregation mechanism we propose is fairly general, depending only on the modification of kinetic rates for protein clearance and it could therefore be relevant also
for other neurodenerative pathologies. For instance, it would be extremely interesting to clarify the possible role of statins in A$\beta$  aggregation and more generally on the development of Alzheimer's disease.

\section*{Methods}

\subsection*{Cell lines and treatments } 
Human Hela cells (ATCC-CRL-2)  or HeLa-GFP cells (AKR-213 Cell Biolabs Inc.) were cultured in 80\%DMEM, 10\% FBS with the addition of 1\% antibiotics and 1\% L-glutamine at 37$^\circ$C / 5\% CO$_2$. SIM (S-6196, Sigma) needs to be activated by opening of the lactone ring before use in cell culture. We used the protocol described by ~\cite{dong2009}. Briefly, 40mg/ml of SIM were dissolved in 100\% ethanol with subsequent addition of 0.1N NaOH ~\cite{dong2009}. The solution was heated at 50$^\circ$C for 2h and then neutralized with HCl to pH 7.2. The resulting solution was bought to a final volume (1ml) with distilled water and aliquots were stored at -80$^\circ$C for no more then 3 months.  Rosuvastatin (SML1264, Sigma) was dissolved in DMSO stock solution (20mM) and stored in aliquots at -20$^\circ$C.  Pravastatin (P4498 Sigma) was dissolved in distilled water and stored in aliquots at -80$^\circ$C. Betulin  (B9757, Sigma) was dissolved in 100\% etanol (500$\mu$M) and stored in aliquots  at -20$^\circ$C.
The medium has been replaced every 4 days with fresh medium containing statins or betulin or fresh medium
for untreated cells. 

\subsection*{Immunofluorescence}   
Subconfluent cells were fixed with 3.7\% paraformaldeide, incubated with 0.1\%Triton X-100 in PBS for 15min at room temperature, then with 3\%BSA in PBS for 30min at room temperature and finally with anti-SREBP-1 (1:100, ab28481 Abcam. 3.7\% PFA fixed cells were incubated with 1\% BSA/10\% goat serum/0.3M glycine/0.1\% tween in PBS for 1h at room temperature and incubated   with anti-rabbit (1:250, Alexa Fluor 488, ab 150077) for 1h at room temperature. 
1$\mu$g/ml Actinstain-555 phalloidin (Cytoskeleton Inc) is used to stain actin for 1h at room temperature.
The nuclei were counterstained with DAPI and the slides mounted with Pro-long anti fade reagent (Life technologies).  The images were acquired with a Leica TCS NT confocal microscope.

\subsection*{Duolink assay}
Subconfluent cells were fixed with 4\% paraformaldeide and incubated with 0.1\%Triton X-100 in PBS for 15min at room temperature, then with 5\%BSA in PBS for 1h at room temperature and finally incubated with anti-neuroserpin  (ab32901 AbCam) coniugated PLA probes (1:50) overnight at 4$^\circ$C.  We detected neuroserpin aggregates with Duolink In Situ staining. The Duolink In Situ staining are based on in situ PLA, which is a proximity ligation assay technology. A pair of oligonucleotide labeled secondary antibodies (PLA probes) generates a signal only when the two PLA probes have bound in close proximity 
(at a distance of less than 40nm), either to the same primary antibody or to two primary antibodies 
bound to the sample in close proximity. The signal from each detected pair of PLA probes is visualized as an individual fluorescent spot. Duolink In Situ Probemaker PLUS enables quick and convenient conjugation of the PLA PLUS oligonucleotide arm directly to the primary antibody (DUO92009 and DUO92010 for PLUS and MINUS probe). The cross reacted proteins were detected according to the manufacturer’s instruction (sigma.com/duolin, section 7.3 Detection protocol). The nuclei were mounted with Duolink in situ mounting medium containing DAPI (DUO82040, Sigma). The images were acquired with a Leica TCS NT confocal microscope. These PLA signals were quantified as described in the section spot detection.

\subsection*{Cholesterol staining} 
Cells were fixed with 4\% paraformaldehyde (PFA) in PB 0.1M for 10 minutes at room temperature (RT). Then, they were incubated with 50$\mu$g/ml Digitonin for 5 minutes and then for 1 hour with 4\% Bovine Serum Albumin (BSA) in PBS. The cells were then incubated with  TNM-AMCA (1$\mu$M, gently gifted by Shinichi Nishimura and Minoru Yoshida, RIKEN Center for Sustainable Resource Science, Saitama, Japan) for 1 hour. All the specimens were mounted with Fluoromount Aqueous Mounting Medium without DAPI (Sigma-F4680) and the images were acquired with  Leica SP2  laser scanning confocal microscope. 

\subsection*{Western Blot} 
Confluent cells were lysed by boiling in a modified Laemmli sample buffer (2\% SDS, 20\% glycerol, and 125 mM Tris-HCl, pH6.8). The protein concentration was measured by BCA Protein Assay Kit (Thermo Scientific). Equal amount of proteins were loaded on gel and separated by SDS-PAGE and transferred to a PVDF membrane (Whatman). After blocking, primary and HRP-linked secondary antibodies, specific bindings were detected by chemiluminescence system (GE Healthcare). Goat polyclonal antibody anti-neuroserpin (1:500, ab32901, AbCAM) or rabbit anti-HMGCR (1:1000, MAB5374, Millipore) or anti-SREPB1 (1:500, ab28481, AbCam) or anti-SREBP2 (1:100, ab30682, AbCam) were used overnight at 4$^\circ$C. Mouse anti-vinculin (1:10000, V9264, Sigma) or anti-$alpha$ Tubulin antibody (1:5000, Sigma) for 1h at room temperature was used as housekeeping. A secondary antibody anti goat-HRP (1:5000, ECL Blotting reagents (GE Healthcare RPN2109) / SuperSignal\texttrademark West Femto Maximum Sensitivity Substrate, Thermo scientific) was used for 1h at room temperature to detect chemiluminiscence. 

For non denaturating gels, cells were resuspended in 50mM Tris-HCl, pH7.4 containing 5mM EDTA, 1\% Triton-X100 and Protease inhibitors cocktail (1ml/100ml lysate, P8340, Sigma), passed for 4 times through an insulin syringe and maintained in ice for 20min. After a brief centrifugation, the protein concentration was measured by BCA Protein Assay Kit (Thermo Scientific). 100$\mu$g protein were loaded on 10\% non denaturating gel, transferred on PVDF membrane as described above. The PVDF membrane was reversible stained with Ponceau S solution (P7170, Sigma) to verify the correct loading of the samples. 

\subsection*{Spot detection}
Image analysis of Duolink In Situ staining was performed using the bioimage informatics platform Icy~\cite{Chaumont2012}. To detect spots, we first use the Thresholder plugin to eliminate the random noise in the images. We used the automated mode so that the algorithm automatically calculates the optimal threshold value. We then applied the Spot Detector plugin~\cite{olivo2002} that is based on an UnDecimated Wavelet Transform tool designed to detect spots.  Then, we used the default values, enabling the scale=2 setup with a threshold equal to 100 and a size filter to accept only spots with a size bigger than three pixels.  We also extracted the number of cell nuclei and compute the ratio between the number of spots and the number of cells.

\subsection*{Statistical Analysis} 
Statistical significance analysis is performed using the Kolmogorov-Smirnov test.

\subsection*{Polymer aggregation model}
Simulations are performed using the protein aggregation model described in Ref. \cite{budrikis2014} where individual protein molecules sit on  a three dimensional cubic lattice. Monomers diffuse with rate $k_{\rm D}$ and attach to neighboring monomers or polymer endpoints with rate $k_{\rm H}$. Polymers move collectively by reptation with a length-dependent rate $k_{\rm R}/i^2$, where $i$ is the number of monomers in the polymer, and locally by end rotations, with rate $k_{\rm E}$ and kink motions with rate $k_{\rm K}$. A polymer can attach to another polymer with rate $k_{\rm H}$ if their endpoints meet, and can fragment by breaking an internal bond with rate $k_{\rm f}$. Inspired by experimental results on neuroserpin polymerization \cite{Noto2012}, we allow for polymerization after at least one of the monomers has been activated with rate $k_{\rm A}$. Active monomers can also become latent with rate $k_{\rm L}$ and after that they do not aggregate. We consider a system of size ($L \times L \times H$) with $L=100 L_0$ and $H=25L_0$, where $L_0$ is the
linear size of a monomer, with periodic boundary conditions along x and y and closed boundary conditions along z.  Monomers enter the system from both closed boundaries with rate $k_{\rm in}$ and monomers and polymers can exit from the same boundaries with rate  
$k_{\rm out}/i^3$. Numerical simulations are performed using Gillespie Montecarlo algorithm \cite{Gillespie1976}.
We measure time in units of  $1/k_{\rm in}$ and set $k_{\rm H}=k_{\rm E}=k_{\rm R}=k_{\rm K}=k_{\rm A}=k_{\rm in}$, $k_{\rm L}=0$, $k_{\rm D}=10^2k_{\rm in}$, $k_{\rm f}=10^{-3}k_{\rm in}$ and vary $k_{\rm out}$.

\subsection*{Vesicle formation model}
We consider a simple model with three basic interaction energies: two bending energies related, respectively, to the elasticity of the membrane and external cage, and a potential representing their mutual interaction. 
The membrane is modeled by a polygon with $N_m$ nodes connected by inextensible bonds and resisting bending through an 
angular spring of energy
\begin{equation}
E_a=K(\theta-\theta_0)^2,
\label{angle_energy}
\end{equation}
where the parameter $\theta_0=\theta_m=0$ is the equilibrium value of the angle between two bonds, and $K=K_m$ is the bending stiffness of the
membrane.  The coating scaffold is modeled, instead, as an open polygon with $N_s$ nodes, rigid bonds and a bending energy given by Eq.\ref{angle_energy} with $K=K_s$ and $\theta_0=\theta_s \geq 0$. In order to simulate a realistic structure, we chose $N_m/N_s=10$, corresponding 
to the ratio between the typical lipid distance in a biological membrane and the characteristic length of the COPII proteins in a cage 
\cite{Zanetti2013}.  For the membrane-scaffold interaction we employ a Lennard-Jones potential where the depth of the potential
well is equal to $\epsilon_{LJ}$. Starting from a pseudo-circular equilibrium configuration, where the scaffold does not perturb the membrane, we simulate the vesicle formation by changing gradually the equilibrium angle of the scaffold $\theta_s$.  The simulations are performed using the LAMMPS molecular dynamics simulator package~\cite{Plimpton1995} with a constant NVE integration and the addition of a viscous damping force to reduce oscillations.


\section*{Acknowledgements:}
We thank S. Nishimura and M. Yoshida, RIKEN Center for Sustainable Resource Science, Saitama, Japan who kindly provided the cholesterol probe TNM-AMCA; GC and SZ are supported by the ERC advanced grant SIZEFFECTS. SZ acknowledges support from the Academy of Finland FiDiPro progam, project 13282993. CAMLP thanks the visiting fellow program of Aalto Science Institute.

\section*{Author contributions:}
CG carried out the native elecrophoresis, MCL and FM carried out immunofluorescences and western blots experiments, 
GC performed numerical simulations and image analysis, SZ designed the model and assisted in the writing of the paper. CAMLP designed the project, performed and supervised experiments, and wrote the paper.

\section*{Additional information}

\textbf{Competing financial interests} The authors declare no competing financial interests.

\end{document}